\pdfoutput=1

\documentclass[11pt]{article}

\usepackage[final]{acl}

\usepackage{times}
\usepackage{latexsym}
\usepackage{multirow}
\usepackage{booktabs}
\usepackage{listings}
\usepackage{makecell}
\usepackage{xcolor}
\usepackage{algorithm}
\usepackage{algpseudocode}  
\usepackage{amsfonts}
\usepackage{pifont}
\newcommand{\cmark}{\ding{51}}%
\newcommand{\xmark}{\ding{55}}%

\definecolor{codegreen}{rgb}{0,0.6,0}
\definecolor{codegray}{rgb}{0.5,0.5,0.5}
\definecolor{codepurple}{rgb}{0.58,0,0.82}
\definecolor{backcolour}{rgb}{0.95,0.95,0.92}

\lstdefinestyle{codestyle}{
    backgroundcolor=\color{backcolour},   
    commentstyle=\color{codegreen},
    keywordstyle=\color{magenta},
    numberstyle=\tiny\color{codegray},
    stringstyle=\color{codepurple},
    basicstyle=\ttfamily\footnotesize,
    breakatwhitespace=false,         
    breaklines=true,                 
    captionpos=b,                    
    keepspaces=true,                 
    numbers=left,                    
    numbersep=5pt,                  
    showspaces=false,                
    showstringspaces=false,
    showtabs=false,                  
    tabsize=2
}
\usepackage[T1]{fontenc}

\usepackage[utf8]{inputenc}

\usepackage{microtype}

\usepackage{inconsolata}

\usepackage{graphicx}

\title{VERSA: A Versatile Evaluation Toolkit for Speech, Audio, and Music}

\author{
 \textbf{Jiatong Shi\textsuperscript{1}},
 \textbf{Hyejin Shim\textsuperscript{1}},
 \textbf{Jinchuan Tian\textsuperscript{1}},
 \textbf{Siddhant Arora\textsuperscript{1}},
 \\
 \textbf{Haibin Wu\textsuperscript{2}},
 \textbf{Darius Petermann\textsuperscript{3}},
 \textbf{Jia Qi Yip\textsuperscript{4}},
 \textbf{You Zhang\textsuperscript{5}},
 \textbf{Yuxun Tang\textsuperscript{6}},
 \\
 \textbf{Wangyou Zhang\textsuperscript{7}},
 \textbf{Dareen Alharthi\textsuperscript{1}},
 \textbf{Yichen Huang\textsuperscript{1}},
 \textbf{Koichi Saito\textsuperscript{8}},
 \textbf{Jionghao Han\textsuperscript{1}},
 \\
 \textbf{Yiwen Zhao\textsuperscript{1}},
 \textbf{Chris Donahue\textsuperscript{1}},
 \textbf{Shinji Watanabe\textsuperscript{1}},
\\
\\
 \textsuperscript{1}Carnegie Mellon University,
 \textsuperscript{2}Microsoft,
 \textsuperscript{3}Indiana University, \\
 \textsuperscript{4}Nanyang Technological University,
 \textsuperscript{5}University of Rochester, \\
 \textsuperscript{6}Renmin University of China, 
 \textsuperscript{7}Shanghai Jiaotong University,
 \textsuperscript{8}Sony AI
}

\begin{document}
\maketitle
\begin{abstract}
In this work, we introduce \textit{VERSA}, a unified and standardized evaluation toolkit designed for various speech, audio, and music signals. The toolkit features a Pythonic interface with flexible configuration and dependency control, making it user-friendly and efficient. With full installation, \textit{VERSA} offers 65 metrics with 729 metric variations based on different configurations. These metrics encompass evaluations utilizing diverse external resources, including matching and non-matching reference audio, text transcriptions, and text captions. As a lightweight yet comprehensive toolkit, \textit{VERSA} is versatile to support the evaluation of a wide range of downstream scenarios. To demonstrate its capabilities, this work highlights example use cases for \textit{VERSA}, including audio coding, speech synthesis, speech enhancement, singing synthesis, and music generation. The toolkit is available at \url{https://github.com/wavlab-speech/versa}.
\end{abstract}

\section{Introduction}
\label{sec:intro}

With the rapid advancements in artificial intelligence-generated content~(AIGC), deep generative models have demonstrated remarkable capabilities in producing high-quality outputs across various domains, including image, video, and sound generation. As generative models become increasingly sophisticated, the need for comprehensive AIGC evaluation has grown, aimed at identifying the strengths and weaknesses of the generated outputs.

As an essential part of the language processing community, diverse generative models for speech, music, and general audio have shown significant potential in applications such as conversational interfaces~\cite{mctear2002spoken}, entertainment~\cite{fraser2018spoken, fancourt2019present, dash2024ai}, and task management~\cite{kulkarni2019conversational}. Due to the perceptual nature of sound-based applications, human subjective assessment is widely regarded as the gold standard for evaluating sound generative models. 

The most fundamental and widely used metric for these models is the mean opinion score~(MOS)~\cite{recommendation1994telephone}. The initial purpose of MOS was to measure the naturalness of generated audio, but it has since evolved into various specific forms, such as evaluating speaker similarity~\cite{toda2016voice}, comparative performance against baseline systems~\cite{harada19995}, emotional similarity~\cite{choi2019multi}, and alignment with prompts~\cite{guo2023prompttts, li24y_interspeech}. Despite its importance, subjective evaluations relying on human input are challenging to conduct due to their labor-intensive nature. Furthermore, achieving statistically significant results often requires a substantial number of samples~\cite{rosenberg2017bias}. Additionally, range-equalizing bias is frequently observed in MOS evaluations due to the psychological grounding of human subjective assessments~\cite{cooper23_interspeech, le2024limits}. Such biases introduce considerable challenges in achieving comparable results across different evaluation datasets and participants, thereby complicating the process of benchmarking generative models effectively. Last but not least, certain evaluation variants may require individuals with expert knowledge to conduct the assessment, particularly in the music domain~\cite{ji2020comprehensive}.

An alternative approach is to develop automatic metrics that align with human preferences. The design of such metrics can vary based on the use of external references and the specific application scenarios, which further complicates their selection.

\begin{figure}
    \centering
    \includegraphics[width=\linewidth]{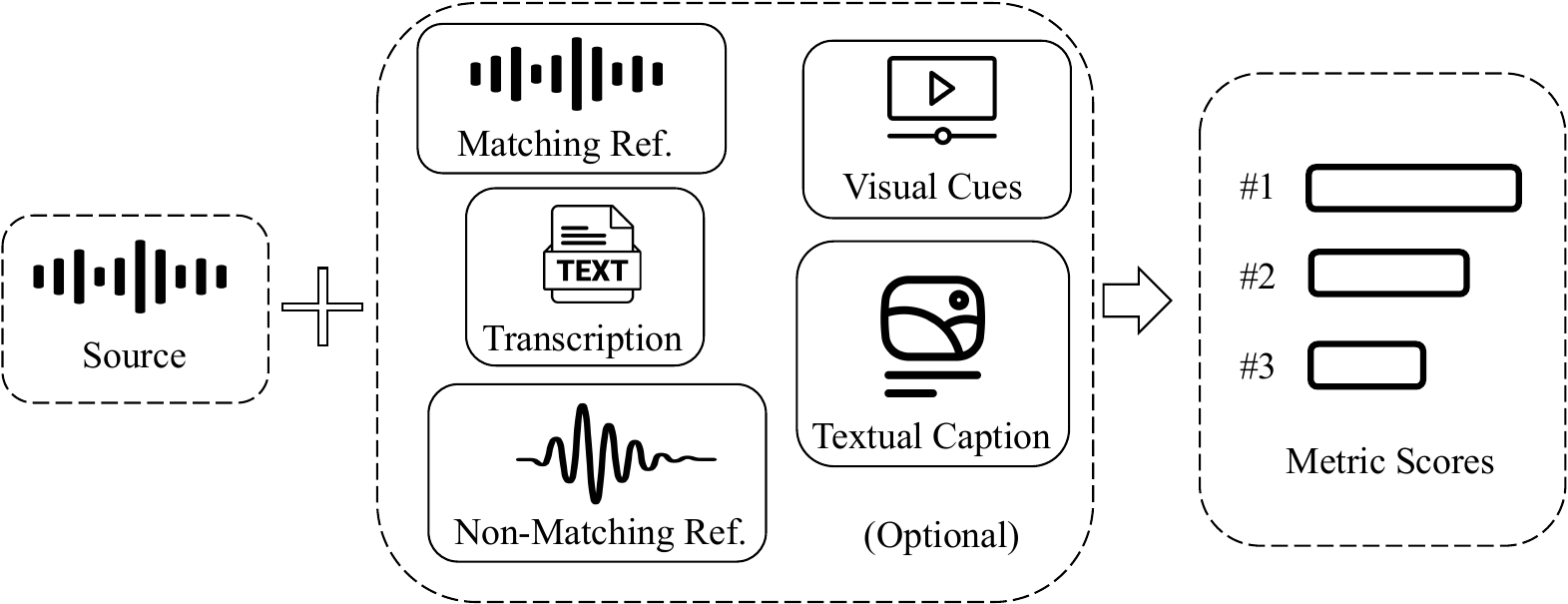}
    \caption{Using various external resources for automatic sound evaluation. External resources include any matching reference signals, non-matching reference signals, transcriptions, visual cues, or textual captions.}
    \label{fig:external-resources}
\end{figure}

The basic setup for evaluation involves using only the candidate audio being evaluated~\cite{liang1994output, falk2006single, lo19_interspeech, saeki22c_interspeech}, which has been actively discussed in recent VoiceMOS challenges~\cite{huang22f_interspeech, cooper2023voicemos, huang2024voicemos}. In addition, a variety of external resources can be optionally utilized for evaluation, as illustrated in Fig.\ref{fig:external-resources}. These resources may include matching reference audio~\cite{rix2001perceptual}, non-matching reference audio~\cite{NORESQA-Manocha2021,ragano2024nomad}, textual transcription~\cite{hayashi2020espnet}, textual captions~\cite{huang2023make}, or visual cues~\cite{hu2021neural}.\footnote{Note that we specifically exclude pre-trained models from our notion of ``external resources''.}

The focus of automatic metrics can vary significantly depending on the application scenario. For instance, in voice conversion, discriminative speaker embeddings can be employed to measure speaker similarity between the generated speech and speech from the same speakers~\cite{das2020predictions}. In cases involving domain-specific content, such as singing, pre-trained models in singing voices may better align with assessments of singing naturalness~\cite{tang2024singmos}. For a fine-grained, sample-level generation, signal-to-noise ratio-related metrics are more suitable, particularly for tasks such as speech enhancement and separation~\cite{luo2019convtasnet}. Due to the creative nature of music, distributional metrics are often better suited for evaluating music generation~\cite{kilgour19_interspeech}. Given the diversity of audio signals, effectively evaluating them requires a broad understanding of general audio events and their contextual relevance~\cite{huang2023make}.

Considering the diversity of metrics and the evolution of generative models for speech, audio, and music, the need for standardized evaluation metrics has become increasingly evident. Without a unified framework, the diversity in evaluation methodologies often leads to inconsistent results, making it challenging to benchmark models or assess advancements effectively. Standardization ensures that all systems are evaluated under comparable conditions, fostering fairness, reproducibility, and meaningful insights across studies. Moreover, centralizing metrics within a single toolkit not only reduces redundancy and inefficiencies but also encourages collaboration by providing researchers with a shared foundation for assessing performance. This need for consistency and centralization underpins the development of \textit{VERSA}, a toolkit designed to address these challenges.

Extended from its prior version in \cite{shi2024espnet}, this work introduces a versatile evaluation toolkit for speech, audio, and music: \textit{VERSA}. Built on a Pythonic interface, \textit{VERSA} integrates 65 metrics and more than 729 variants, offering a wide array of automatic evaluation tools tailored for speech, audio, and music. By providing diverse metrics, \textit{VERSA} aims to serve as a one-stop solution for multi-domain, multi-level, and multi-focus sound evaluation across various downstream applications. With examples showcasing the use of \textit{VERSA} in \textit{ESPnet-Codec} and \textit{ESPnet-SpeechLM}, we anticipate that the \textit{VERSA} toolkit will become a key component in advancing sound generation benchmarks, addressing challenges in speech generation, and supporting multi-modal generative frameworks. The codebase is publicly available at \url{https://github.com/wavlab-speech/versa}.\footnote{A video demonstration is available at \url{https://youtu.be/t7UP1uFvaCM}.}

\section{VERSA}

This section details the design of \textit{VERSA}, focusing on its general framework, supported metrics, and potential benefits for the community.

\subsection{VERSA Framework}
\label{sec:versa framework}

\begin{figure}
    \centering
    \includegraphics[width=1.1\linewidth]{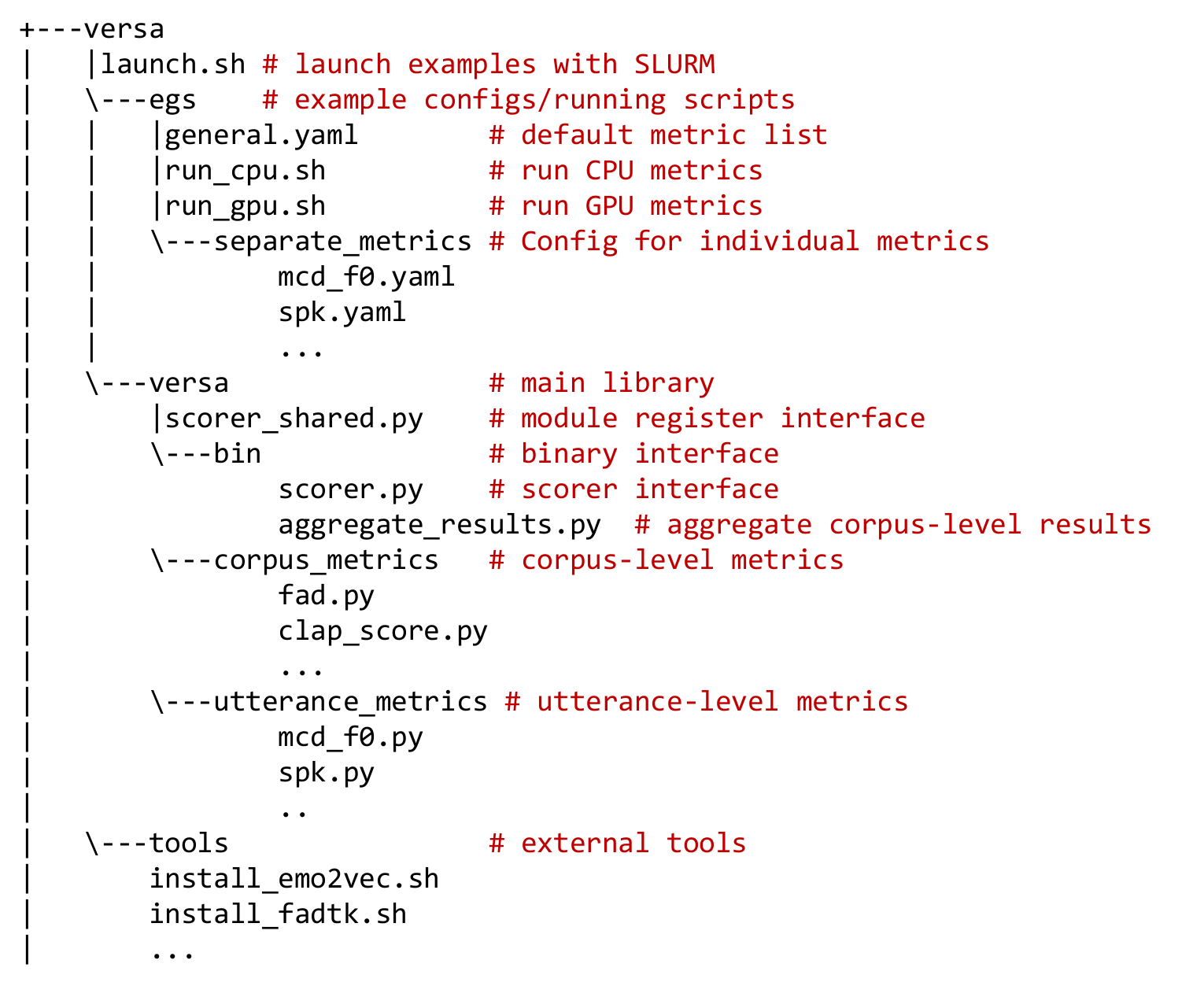}
    \caption{Directory structure of \textit{VERSA}. Detailed discussion can be found in Sec.~\ref{sec:versa framework}}
    \label{fig:versa-dir}
\end{figure}

As illustrated in Fig.~\ref{fig:versa-dir}, the core library of \textit{VERSA} is implemented as a Python package with two straightforward interfaces: \texttt{scorer.py} and \texttt{aggregate\_result.py}. The \texttt{scorer.py} interface computes the automatic metrics, while \texttt{aggregate\_result.py} consolidates the results into a final report for users.

Once installed via \texttt{pip}, using \textit{VERSA} is as simple as the following:

\begin{lstlisting}[language=, label=list-versa ,caption=A simple scorer interface.]
python versa/bin/scorer.py \
    --score_config egs/general.yaml \
    --gt <ground truth audio list> \
    --pred <candidate audio list> \
    --output_file test_result
\end{lstlisting}
where the ground truth audio list is optional and is not required for independent metrics.

\noindent \textbf{I/O Interface}: \textit{VERSA} offers three I/O interfaces for handling audio samples: the \texttt{Soundfile} interface, the \texttt{Directory} interface, and the \texttt{Kaldi} interface. These interfaces support a variety of audio formats (e.g., PCM, FLAC, MP3, Kaldi-ARK) and different file organizations, such as \texttt{wav.scp} files or individual audio files stored within a parent directory. For each (candidate, reference) audio signal pair, a resampling is conducted for each metric according to the specific sampling rate required by that metric. \texttt{librosa} is used for resampling.


\noindent \textbf{Flexible Configuration}: \textit{VERSA} employs a unified \texttt{YAML}-style configuration file to define which metrics to use and control their detailed setups. While users can directly explore the library code, we also provide example configuration files for different metrics in the \texttt{egs} directory, as shown in Fig.~\ref{fig:versa-dir}. For instance, \texttt{egs/general.yaml} provides a configuration template for the default installation metrics. Additionally, individual \texttt{YAML} configuration files for specific metrics are available under \texttt{egs/separate\_metric}. Some example configurations are discussed in Appendix~\ref{sec:config}.

\noindent \textbf{Strict Dependency Control}: Managing dependencies can be challenging when using diverse evaluation metrics. To address this, \textit{VERSA} offers a minimal-dependency installation that supports a core set of metrics by default, while additional installation scripts are provided for metrics with extra requirements. This approach significantly reduces dependency overhead during \textit{VERSA} installation, especially for metrics with heavy dependencies or complex compilation needs. As shown in Fig.~\ref{fig:versa-dir}, these additional installation scripts are located in the \texttt{tools} directory.

To ensure correct model usage, many official packages released by model providers enforce strict dependency control by specifying exact package versions (e.g., specific versions of PyTorch or NumPy). While this ensures compatibility, it often introduces unnecessary dependency conflicts with major packages. To provide a more flexible environment for users, \textit{VERSA} bypasses these strict version requirements by adapting the interfaces of such metrics into our own local forks. These forks are supplemented with additional numerical tests to ensure functionality without adhering to rigid version control.

Moreover, using our own fork allows us to integrate \textit{VERSA}-specific interfaces into external metrics that may otherwise conflict with the toolkit’s design philosophy. This flexibility ensures a consistent and seamless interface across the \textit{VERSA} library, enhancing usability and maintaining the toolkit's design integrity.

In Appendix~\ref{sec: add-system-design}, we further discuss additional system design concepts on job scheduling, test protocol, resource management, and community-driven contribution guidelines.

\begin{table*}[]
    \centering
        \caption{List of supported metrics in \textit{VERSA}. The ``Base" column indicates whether the metrics are included in the minimum installation of VERSA. The ``Model Based'' column represents metrics that need pre-trained models. The ``Target Direction'' column indicates which direction is desirable for each metric without being overly technical.}
    \resizebox{\linewidth}{!}{
    \begin{tabular}{c|c|c|c|c|l|c|c|c|c|c|c}
    \toprule
       \multirow{2}{*}{No.} & \multirow{2}{*}{Type} & \multicolumn{3}{|c|}{Domain} & \multirow{2}{*}{Name}  & \multirow{2}{*}{Base} & \multirow{2}{*}{Variants} & \multirow{2}{*}{Range} & \multirow{2}{*}{\makecell{Model\\Based}} & \multirow{2}{*}{\makecell{Target\\Direction}} & \multirow{2}{*}{Reference} \\
       \cmidrule{3-5}
       & & Speech & Audio & Music & & & &  &  & \\
       \midrule
        1 & \multirow{24}{*}{Independent} & \cmark & \xmark & \xmark & Deep Noise Suppression MOS Score of P.835 (DNSMOS P.835)  & \cmark & 1 & [1, 5] & \cmark & $\uparrow$ & \small \cite{reddy2022dnsmos} \\
        2 &  & \cmark & \xmark & \xmark & Deep Noise Suppression MOS Score of P.808 (DNSMOS P.808)  & \cmark & 1 & [1, 5] & \cmark & $\uparrow$ & \small \cite{reddy2021dnsmos} \\
        3 &  & \cmark & \xmark & \xmark & Speech Quality and Naturalness Assessment (NISQA) & \cmark & \href{https://github.com/gabrielmittag/NISQA?tab=readme-ov-file\#using-nisqa}{3} & [1, 5] & \cmark & $\uparrow$ & \small \cite{mittag21_interspeech} \\
        4 &  & \cmark & \xmark & \xmark & UTokyo-SaruLab System for VoiceMOS 2022 (UTMOS) & \cmark & 1 & [1, 5] & \cmark & $\uparrow$ & \small \cite{saeki22c_interspeech} \\
        5 &  & \cmark & \cmark & \cmark & Packet Loss Concealment-focus MOS (PLCMOS) & \cmark & 1 & [1, 5] & \cmark & $\uparrow$ & \small \cite{diener23_interspeech} \\
        6 &  & \cmark & \xmark & \xmark & Torch-Squim PESQ (TS-PESQ) & \cmark & 1 & [1, 5] & \cmark & $\uparrow$ & \small \cite{kumar2023torchaudio} \\
        7 &  & \cmark & \xmark & \xmark & Torch-Squim STOI (TS-STOI) & \cmark & 1 & [0, 1] & \cmark & $\uparrow$ & \small \cite{kumar2023torchaudio} \\
        8 &  & \cmark & \xmark & \xmark & Torch-Squim SI-SNR (TS-SI-SNR) & \cmark & 1 & (-inf, inf) & \cmark & $\uparrow$ & \small \cite{kumar2023torchaudio} \\
        9 &  & \xmark & \xmark & \cmark & Singing voice MOS (SingMOS) & \cmark & 1 & [1, 5] & \cmark & $\uparrow$ & \small \cite{tang2024singmos} \\
        10 &  & \cmark & \xmark & \cmark & Subjective Speech Quality Assessment (SSQA) in SHEET Toolkit & \cmark & 1 & [1, 5] & \cmark & $\uparrow$ & \small \cite{huang2024mos} \\
        11 &  & \cmark & \xmark & \xmark & UTokyo-SaruLab System for VoiceMOS 2024 (UTMOSv2) & \xmark & 1 & [1, 5] & \cmark & $\uparrow$ & \small \cite{baba2024t05} \\
        12 &  & \cmark & \xmark & \xmark & Speech Quality with Contrastive Regression (SCOREQ) wo. Ref. & \xmark & \href{https://github.com/alessandroragano/scoreq?tab=readme-ov-file\#using-scoreq-from-the-command-line}{2} & [1, 5]  & \cmark & $\uparrow$ & \small \cite{ragano2024scoreq} \\
        13 &  & \cmark & \xmark & \xmark & Speech Enhancement-based SI-SNR (SE-SI-SNR) & \cmark & \href{https://huggingface.co/models?pipeline_tag=audio-to-audio\&library=espnet}{69} & (-inf, inf) & \cmark & $\uparrow$ & \small \cite{zhang2024urgent} \\
        14 &  & \cmark & \xmark & \xmark & Speech Enhancement-based CI-SDR (SE-CI-SDR) & \cmark  & \href{https://huggingface.co/models?pipeline_tag=audio-to-audio\&library=espnet}{69} & (-inf, inf) & \cmark & $\uparrow$ & \small \cite{zhang2024urgent} \\
        15 &  & \cmark & \xmark & \xmark & Speech Enhancement-based SAR (SE-SAR) & \cmark  & \href{https://huggingface.co/models?pipeline_tag=audio-to-audio\&library=espnet}{69} & (-inf, inf) & \cmark & $\uparrow$ & \small \cite{zhang2024urgent} \\
        16 &  & \cmark & \xmark & \xmark & Speech Enhancement-based SDR (SE-SDR) & \cmark  & \href{https://huggingface.co/models?pipeline_tag=audio-to-audio\&library=espnet}{69} & (-inf, inf) & \cmark  & $\uparrow$ & \small \cite{zhang2024urgent} \\
        17 &  & \xmark & \cmark & \xmark & Prompting Audio-Language Models (PAM) metric & \cmark & 1 & [0, 1] & \cmark & $\uparrow$ &   \small \cite{deshmukh2024pam} \\
        18 &  & \cmark & \cmark & \xmark &  Speech-to-reverberation Modulation Energy Ratio (SRMR) & \xmark & 1 & [0, inf) & \cmark & $\uparrow$ & \small \cite{falk2010non} \\
        19 &  & \cmark & \xmark & \xmark & Voice Activity Detection (VAD) & \cmark & 1 & -  & \cmark & - &\small \cite{Silero-VAD} \\
        20 &  & \cmark & \xmark & \xmark & Speaker Turn Taking (SPK-TT) & \xmark & 1 & [0, inf) & \cmark & - & \small \cite{goodwin1990conversation} \\
        21 &  & \cmark & \xmark & \xmark & Speaking Word Rate (SWR) & \cmark & \href{https://github.com/openai/whisper?tab=readme-ov-file\#available-models-and-languages}{7} & [0, inf) & \xmark & - & \small \cite{radford2023robust} \\
        22 &  & \cmark & \xmark & \xmark & Anti-spoofing Score (SpoofS) & \cmark & \href{https://github.com/clovaai/aasist/tree/main/models/weights}{2} & [0, 1] & \cmark & $\uparrow$ & \small \cite{Jung2021AASIST} \\
        23 &  & \cmark & \xmark & \xmark & Language Identification (LID) & \cmark & \href{https://huggingface.co/models?pipeline_tag=automatic-speech-recognition\&library=espnet\&search=owsm}{17} & - & \cmark & - & \small \cite{peng2023reproducing} \\
        24 & & \cmark & \cmark & \cmark & Audiobox Aesthetics (AA) & \xmark & 1 & [1, 10] & \cmark & $\uparrow$ & \small \cite{tjandra2025meta} \\
        \midrule
        25 & \multirow{28}{*}{Dependent} & \cmark & \cmark & \cmark &  Mel Cepstral Distortion (MCD)  & \cmark & 1 & [0, inf) & \xmark & $\downarrow$ & \small \cite{kubichek1993mel} \\
        26 &  & \cmark & \xmark & \xmark &  F0 Correlation (F0-CORR) & \cmark & 1 & [-1, 1] & \xmark & $\uparrow$ & \small \cite{hayashi2021espnet2} \\
        27 &  & \cmark & \xmark & \xmark &  F0 Root Mean Square Error (F0-RMSE) & \cmark & 1 & [0, inf) & \xmark & $\downarrow$ & \small \cite{hayashi2021espnet2} \\
        28 &  & \cmark & \cmark & \cmark &  Signal-to-interference Ratio (SIR) & \cmark & 1 & (-inf, inf) & \xmark& $\uparrow$ & \small \cite{fevotte2005bss_eval} \\
        29 &  & \cmark & \cmark & \cmark &  Signal-to-artifact Ratio (SAR) & \cmark & 1 &  (-inf, inf) & \xmark & $\uparrow$ & \small \cite{fevotte2005bss_eval} \\
        30 &  & \cmark & \cmark & \cmark &  Signal-to-distortion Ratio (SDR) & \cmark & 1 & (-inf, inf) & \xmark & $\uparrow$ & \small \cite{fevotte2005bss_eval} \\
        31 &  & \cmark & \cmark & \cmark &  Convolutional scale-invariant signal-to-distortion ratio (CI-SDR) & \cmark & 1 & (-inf, inf) & \xmark & $\uparrow$ & \small \cite{boeddeker2021convolutive} \\
        32 &  & \cmark & \cmark & \cmark &  Scale-invariant signal-to-noise ratio (SI-SNR) & \cmark & 1 & (-inf, inf) & \xmark & $\uparrow$ & \small \cite{luo2018tasnet} \\
        33 &  & \cmark & \xmark & \xmark & Perceptual Evaluation of Speech Quality (PESQ) & \cmark & 1 & [1, 5] & \cmark & $\uparrow$ & \small \cite{rix2001perceptual} \\
        34 &  & \cmark & \xmark & \xmark &  Short-Time Objective Intelligibility (STOI) & \cmark & 1 & [0, 1] & \xmark & $\uparrow$ & \small \cite{taal2011algorithm} \\
        35 &  & \cmark & \xmark & \xmark & Speech BERT Score (D-BERT) & \cmark & 1 & [-1, 1] & \cmark & $\uparrow$ & \small \cite{saeki24_interspeech} \\
        36 &  & \cmark & \xmark & \xmark & Discrete Speech BLEU Score (D-BLEU) & \cmark & 1 & [0, 1] & \cmark & $\uparrow$ & \small \cite{saeki24_interspeech} \\
        37 &  & \cmark & \xmark & \xmark & Discrete Speech Token Edit Distance (D-Distance) & \cmark & 1 & [0, 1] & \cmark & $\uparrow$ & \small \cite{saeki24_interspeech} \\
        38 &  & \cmark & \cmark & \cmark &  Dynamic Time Warping Cost (WARP-Q) & \xmark & 1 & [0, inf) & \cmark & $\uparrow$ & \small \cite{Wissam_IET_Signal_Process2022} \\
        39 &  & \cmark & \xmark & \xmark & Speech Quality with Contrastive Regression (SCOREQ) w. Ref. & \xmark & \href{https://github.com/alessandroragano/scoreq?tab=readme-ov-file\#using-scoreq-from-the-command-line}{2} & [1, 5] & \cmark & $\uparrow$ & \small \cite{ragano2024scoreq} \\
        40 &  & \cmark & \cmark & \cmark & 2f-Model & \xmark & 1 & [0, 100] & \cmark & $\uparrow$ & \small \cite{8937179} \\
        41 &  & \cmark & \cmark & \cmark &  Log-weighted Mean Square Error (Log-WMSE) & \cmark & 1 & (-inf, inf) & \xmark & $\uparrow$ & \small \cite{logwmse} \\
        42 &  & \cmark & \xmark & \xmark & ASR-oriented Mismatch Error Rate (ASR-Mismatch) & \cmark & \href{https://github.com/openai/whisper?tab=readme-ov-file\#available-models-and-languages}{7} & [0, inf] & \cmark & $\downarrow$ & \small \cite{radford2023robust} \\
        43 &  & \cmark & \cmark & \cmark & Virtual Speech Quality Objective Listener (VISQOL) & \xmark & \href{https://github.com/google/visqol/tree/master/model}{4} &  [1,5] & \cmark & $\uparrow$ & \small \cite{chinen2020visqol} \\
        44 &  & \cmark & \cmark & \cmark &  Frequency-Weighted SEGmental SNR (FWSEGSNR) & \cmark & 1 & (-inf, inf) & \xmark & $\uparrow$ & \small \cite{tribolet1978study} \\
        45 &  & \cmark & \xmark & \xmark &  Weighted Spectral Slope (WSS) & \xmark & 1 & [0, inf)  & \xmark & $\downarrow$ &\small \cite{klatt1982prediction} \\
        46 &  & \cmark & \xmark & \xmark &  Cepstrum Distance (CD) & \xmark & 1 & [0, inf)  & \xmark & $\downarrow$ &\small \cite{barnwell1988objective} \\
        47 &  & \cmark & \xmark & \xmark &  Composite Objective Speech Quality (Csig, Cbak, Covl) & \xmark & 1 & [1, 5]  & \cmark & $\uparrow$ & \small \cite{hu2007evaluation} \\
        48 &  & \cmark & \xmark & \xmark &  Coherence and Speech Intelligibility Index (CSII)& \xmark & 1  & [0, 1] & \xmark & $\uparrow$ & \small \cite{kates2005coherence} \\
        49 &  & \cmark & \xmark & \xmark &  Normalized-Covariance Measure (NCM) & \xmark & 1 & [-1, 1] & \xmark & $\uparrow$  & \small \cite{chen2010analysis} \\
        \midrule
        50 & \multirow{9}{*}{Non-match} &\cmark & \xmark & \xmark &  Non-matching Reference Speech Quality Assessment (Noresqa) & \xmark & 2 & [1, 5] & \cmark & $\uparrow$ & \small \cite{NORESQA-Manocha2021} \\
        51 &  & \cmark & \xmark & \xmark & Torch-Squim MOS (TS-MOS) & \cmark & 1 & [1, 5] & \cmark & $\uparrow$ & \small \cite{kumar2023torchaudio} \\
        52 &  & \cmark & \xmark & \xmark & ESPnet ASR Model Word Error Rate (ESPnet-WER) & \cmark  & \href{https://huggingface.co/models?pipeline_tag=automatic-speech-recognition\&library=espnet}{270} & [0, inf) & \cmark & $\downarrow$ & \small \cite{watanabe18_interspeech} \\
        53 &  & \cmark & \xmark & \xmark & ESPnet OWSM Model Word Error Rate (OWSM-WER) & \cmark & \href{https://huggingface.co/models?pipeline_tag=automatic-speech-recognition\&library=espnet\&search=owsm}{17} & [0, inf) & \cmark & $\downarrow$ & \small \cite{peng2023reproducing} \\
        54 &  & \cmark & \xmark & \xmark & OpenAI Whisper Model Word Error Rate (Whisper-WER) & \cmark & \href{https://github.com/openai/whisper?tab=readme-ov-file\#available-models-and-languages}{7} & [0, inf) & \cmark & $\downarrow$ & \small \cite{radford2023robust} \\
        55 &  & \cmark & \xmark & \xmark & Emotion Similarity (EMO-SIM) & \xmark & 1 & [-1, 1] & \cmark & $\uparrow$ & \small \cite{ma-etal-2024-emotion2vec} \\
        56 &  & \cmark & \xmark & \xmark & Speaker Similarity (SPK-SIM) & \cmark & \href{https://huggingface.co/models?other=speaker-recognition\&search=espnet}{15} & [-1, 1] & \cmark & $\uparrow$ & \small \cite{jung24c_interspeech} \\
        57 &  & \cmark & \xmark & \xmark & Non-Matching Reference Audio Quality Assessment (NOMAD) & \xmark & 1 & [1, 5] & \cmark & $\uparrow$ & \small \cite{ragano2024nomad} \\
        58 &  & \xmark & \cmark & \cmark & Contrastive Language-Audio Pretraining Score (CLAP Score) & \xmark & 1 & [-1, 1] & \cmark & $\uparrow$ &  \small \cite{huang2023make} \\
        59 &  & \xmark & \cmark & \cmark & Accompaniment Prompt Adherence (APA) & \xmark & 1 & [-1, 1] & \cmark & $\uparrow$ &  \small \cite{grachten2024measuring} \\
        60 &  & \cmark & \cmark & \cmark &  Log Likelihood Ratio (LLR) & \xmark & 1 & [0, inf) & \xmark & $\uparrow$ & \small \cite{hu2007evaluation} \\
        \midrule
        61 & \multirow{5}{*}{Distributional} & \xmark & \cmark & \cmark & Fr\'echet Audio Distance Audio Distance (FAD) & \xmark & \href{https://github.com/microsoft/fadtk?tab=readme-ov-file\#supported-models}{11} & [0, inf) & \cmark & $\downarrow$ & \small \cite{48813} \\
        62 &  & \xmark & \cmark & \cmark & Kullback-Leibler Divergence on Embedding Distribution (KLD) & \xmark & \href{https://github.com/microsoft/fadtk?tab=readme-ov-file\#supported-models}{11} & [0, inf) & \cmark & $\downarrow$ &  \small \cite{48813} \\
        63 &  & \xmark & \cmark & \cmark & Density in Embedding Space (Density-Embedding)  & \xmark & \href{https://github.com/microsoft/fadtk?tab=readme-ov-file\#supported-models}{11} & [0, inf) & \cmark & $\uparrow$ & \small \cite{naeem2020reliable} \\
        64 &  & \xmark & \cmark & \cmark & Coverage in Embedding Space (Coverage-Embedding)  & \xmark & \href{https://github.com/microsoft/fadtk?tab=readme-ov-file\#supported-models}{11} & [0, 1] & \cmark & $\uparrow$ & \small \cite{naeem2020reliable} \\
        65 &  & \xmark & \cmark & \cmark & Kernel Distance/Maximum Mean Discrepancy (KID)   & \xmark & \href{https://github.com/microsoft/fadtk?tab=readme-ov-file\#supported-models}{11} & [0, inf) & \cmark & $\downarrow$ & \small \cite{48813} \\
        \midrule
        - & Total & 56 & 23 & 23 & - & 40 & 729 & - & 48 & - \\
        \bottomrule
    \end{tabular}}

    \label{tab:versa-metrics}
\end{table*}
\begin{table*}[]
    \centering
    \caption{Comparison to related toolkits. The number of metrics are collected at 12/09/2024.}
    \resizebox{\linewidth}{!}{
    \begin{tabular}{l|c|c|c|l|c}
    \toprule
       \multirow{2}{*}{Name} & \multicolumn{3}{|c|}{Domain} & \multirow{2}{*}{Open-source Link}  & \multirow{2}{*}{Metric Type}  \\
       \cmidrule{2-4}
       & Speech & Audio & Music & &  \\
    \midrule
    \textit{ESPnet} \cite{watanabe18_interspeech} & \cmark & \xmark & \cmark & \url{https://github.com/espnet/espnet} & 16 \\ 
    \textit{Apmhion} \cite{zhang2023amphion} & \cmark & \cmark & \cmark & \url{https://github.com/open-mmlab/Amphion} & 15 \\ 
    \textit{SHEET} \cite{huang2024mos} & \cmark & \xmark & \cmark & \url{https://github.com/unilight/sheet} & 3 \\ 
    \textit{SpeechMOS} & \cmark & \cmark & \cmark & \url{https://pypi.org/project/speechmos} & 3 \\
    \textit{BSS-EVAL} \cite{fevotte2005bss_eval} & \cmark & \cmark & \cmark & \url{https://pypi.org/project/fast-bss-eval} & 4 \\ 
    \textit{ClearerVoice-SpeechScore} & \cmark & \xmark & \xmark & \url{https://github.com/modelscope/ClearerVoice-Studio} & 14 \\
    \textit{AudioLDM-Eval} \cite{liu2023audioldm} & \xmark & \cmark & \cmark & \url{https://github.com/haoheliu/audioldm_eval} & 9 \\
    \textit{Stable-Audio-Metric} & \cmark & \cmark & \cmark & \url{https://github.com/Stability-AI/stable-audio-metrics} & 3 \\
    \textit{Sony Audio-Metrics} \cite{grachten2024measuring} & \cmark & \cmark & \cmark & \url{https://github.com/SonyCSLParis/audio-metrics} & 4 \\
    \textit{FADTK} \cite{gui2024adapting} & \cmark & \cmark & \cmark & \url{https://github.com/microsoft/fadtk} & 11 \\
    \textit{Pysepm} & \cmark & \xmark & \xmark & \url{https://github.com/schmiph2/pysepm} & 10 \\
    \textit{AudioCraft}~\cite{copet2024simple} & \xmark & \cmark & \cmark & \url{https://github.com/facebookresearch/audiocraft/blob/main/docs/METRICS.md} & 6 \\
    \textit{AutaTK} \cite{vinay2023aquatk} & \xmark & \cmark & \cmark & \url{https://github.com/Ashvala/AQUA-Tk} & 9 \\
    \midrule
    \textit{VERSA} & \cmark & \cmark & \cmark & \url{https://github.com/shinjiwlab/versa} & 65 \\
       \bottomrule
    \end{tabular}}

    \label{tab:comparison}
\end{table*}

\subsection{Supported Metrics}
\label{ssec:supported metrics}

\textit{VERSA} stands out for its extensive range of supported metrics, categorized into four main types:

\noindent \textbf{Independent metrics}: These metrics do not require any dependent external resources, other than pre-trained models. 
Notably, we adopt an extended form of independent metrics in speech and audio assessment, which also considers profiling-related metrics, such as voice activity detection and speaker turn-taking. 
\textit{VERSA} currently supports 22  independent metrics.

\noindent \textbf{Dependent metrics}: These metrics rely on matching sound references. In \textit{VERSA}, 25 dependent metrics are supported.

\noindent\textbf{Non-matching metrics}: These metrics use non-matching reference data or different modalities. \textit{VERSA} currently supports 11 non-matching metrics.

\noindent \textbf{Distributional metrics}: These metrics conduct distribution comparisons between two datasets, providing a more general view of the generative models' performance. \textit{VERSA} currently supports 5 distributional metrics.

As summarized in Table~\ref{tab:versa-metrics}, \textit{VERSA} supports a total of 65 metrics, 39 of which are included in the minimal installation. Of these, 54 metrics are applicable to speech tasks, 22 to general audio tasks, and 22 to music tasks. Additionally, several metrics feature variations based on different pre-trained models, such as word error rate, speaker similarity, and Fréchet Audio Distance (FAD) scores. By simply modifying the configuration file,\footnote{An example of a configuration change is provided in Appendix~\ref{ssec:config-change}.} \textit{VERSA} can generate up to 729 distinct metric variants, offering flexibility for a wide range of evaluation scenarios.

\subsection{Advantages of \textit{VERSA}}

This section highlights the key benefits of \textit{VERSA}, focusing on its ability to ensure consistency, facilitate comparability, provide comprehensive insights, and enhance efficiency.

\noindent \textbf{Consistency}: \textit{VERSA} ensures uniform evaluation criteria across experiments, addressing a critical need in the field of sound generative models. By standardizing the implementation of metrics, \textit{VERSA} minimizes the variability introduced by subjective judgments or inconsistent evaluation setups (e.g., coding environments).

\noindent \textbf{Comparability}: One of \textit{VERSA}'s key advantages is its ability to facilitate benchmarking against existing models and methodologies. By providing a unified set of metrics, it ensures that evaluations are conducted under fair and objective conditions. This comparability is important for assessing the relative performance of new approaches, fostering innovation, and enabling the broader research community to progress effectively.

\noindent \textbf{Comprehensiveness}: \textit{VERSA} supports a wide array of evaluation metrics, including dimensions such as perceptual quality, intelligibility, affective information, and creative diversity. By incorporating these diverse measures, the toolkit provides a holistic view of system performance, especially for researchers to gain deeper insights into both the strengths and weaknesses of each method.

\noindent \textbf{Efficiency}: With its all-in-one design, \textit{VERSA} significantly enhances efficiency by supporting multiple metrics within a single toolkit. Users no longer need to rely on separate tools or perform manual calculations to assess different aspects of audio performance. This workflow reduces the time, effort, and potential errors associated with using fragmented evaluation methods, accelerating the overall research and development process.

\section{Comparison to Other Toolkits}

As discussed in Sec.~\ref{sec:intro}, the challenges associated with subjective evaluation have propelled the community toward exploring objective evaluation metrics. The growing demand for sound evaluation toolkits has led to numerous efforts in domain-specific evaluation for sound generation.

In the speech domain, prior to the deep learning era, the International Telecommunication Union Telecommunication Standardization Sector (ITU-T) played a pivotal role in designing evaluation metrics for speech processing tasks such as speech coding and speech enhancement. More recently, text-to-speech (TTS) toolkits have begun incorporating speech quality assessment features, exemplified by \textit{ESPnet-TTS}~\cite{hayashi2020espnet, hayashi2021espnet2} and \textit{Amphion}~\cite{zhang2023amphion}. Additionally, the \textit{Speech Human Evaluation Estimation Toolkit~(SHEET)} framework provides an all-in-one recipe-style toolkit for data preparation, speech quality prediction model training, and evaluation~\cite{huang2024mos}. In speech enhancement, foundational signal-level metrics were first consolidated in~\cite{fevotte2005bss_eval}, followed by extensions such as \textit{Pysepm}, which supports 10 metrics~\cite{loizou2013se}. More recent advancements include the addition of 14 speech enhancement metrics in \textit{ClearerVoice}~\cite{clearvoice}.

In the audio domain, \textit{AudioLDM-Eval} focuses on evaluating audio language models with nine types of metrics~\cite{liu2023audioldm, audioldm2-2024taslp}. Stability AI has introduced three audio metrics~\cite{stable-audio}, while Sony CSL has open-sourced four additional types of audio metrics~\cite{grachten2024measuring}.

In the music domain, \textit{MIR\_EVAL} is a pioneering toolkit that aggregates metrics for music information retrieval tasks~\cite{raffel2014mir_eval}. More recently, Microsoft released \textit{FADTK}, which emphasizes a comprehensive FAD-embedding space for generative music evaluation~\cite{gui2024adapting}.

A summary of the related toolkits is presented in Table~\ref{tab:comparison}. Each framework has made significant contributions to the community, many serving as foundational tools for sound generation research. Building on these previous toolkits, \textit{VERSA} distinguishes itself with a general design applicable across multiple domains and its comprehensive inclusion of 65 metrics with 729 variants—capabilities that have not been achieved before.
\section{Demonstration}
\label{sec:demo}

We demonstrate several use cases of \textit{VERSA} in diverse scenarios, including speech coding, speech synthesis, speech enhancement, singing synthesis, and music generation. 

\noindent \textbf{Speech coding} remains one of the most widely utilized applications within the speech-processing community. In this demonstration, we leverage \textit{VERSA} to evaluate nine publicly available codecs. More details and corresponding results are discussed in Table~\ref{tab:codec} in Appendix~\ref{ssec:codec}.

\noindent \textbf{Text-to-speech} aims to convert text into speech signals. In this demonstration, we use \textit{VERSA} to compare nine open-source TTS models. More details and corresponding are shown in Table~\ref{tab:tts}, Appendix~\ref{ssec:tts}.



\noindent \textbf{Speech enhancement} targets the improvement of speech transmission in different environments. In this paper, we also demonstrate the \textit{VERSA} usage in speech enhancement in three speech enhancement models.
More details and corresponding information are shown in Table~\ref{tab:se}, Appendix~\ref{ssec:se}.

\noindent \textbf{Singing synthesis} is an intersection between speech and music generation, where both speech-oriented and music-oriented evaluation metrics are needed. \textit{VERSA} offers a one-stop solution to this problem based on the collection of a variety of metrics in each domain. In this work, we demonstrate the evaluation of a range of singing voice synthesis models. Table~\ref{tab:svs}, Appendix~\ref{ssec:svs} shows more details and corresponding information.

\noindent \textbf{Music generation} and its evaluation have received increasing attention due to the rapid progress in model development. To consider both creativity and musical harmony, recent evaluation methods mostly utilize distribution metrics for the evaluation, exhibiting a large difference to other sound generation systems. In this demonstration, \textit{VERSA} aggregates a range of music generation evaluation metrics into a single toolkit. More details and corresponding information are shown in Table~\ref{tab:music}, Appendix~\ref{ssec:music}.

\section{Conclusion} 

\label{sec:conclusion}

In this work, we introduced \textit{VERSA}, a comprehensive and versatile evaluation toolkit for assessing speech, audio, and music signals. With its flexible Pythonic interface and extensive suite of metrics, \textit{VERSA} empowers researchers and developers to conduct rigorous and reproducible evaluations across diverse generative tasks. 

Through its integration of more than 65 metrics and 729 variants, \textit{VERSA} provides unparalleled support for evaluating speech synthesis, audio coding, music generation, and more. The toolkit not only simplifies the evaluation process but also addresses challenges such as bias in subjective evaluations and the need for domain-specific expertise.

\section{Ethics Statement}
\textit{VERSA} is designed to address the challenges of evaluating sound generative models in diverse linguistic, cultural, and acoustic contexts. Generative sound models often risk perpetuating biases due to their reliance on datasets that may underrepresent certain languages, accents, and cultural expressions. To mitigate these risks, \textit{VERSA} incorporates evaluation metrics and methodologies that accommodate a wide range of audio characteristics, including tonal, phonetic, and rhythmic variations present across global languages and music traditions.

The toolkit encourages the use of regionally diverse datasets and provides flexibility to integrate culturally specific evaluation resources, such as non-standard phonetic systems or traditional musical structures. By doing so, \textit{VERSA} seeks to foster the development of sound generative models that respect and represent the full spectrum of human audio diversity.

Through these efforts, \textit{VERSA} empowers researchers and developers to create more inclusive generative models, ensuring that advances in speech, audio, and music technologies benefit communities worldwide, regardless of linguistic or cultural background.

\section{Broader Impact}

\subsection{Current Limitations}
While \textit{VERSA} offers a comprehensive and versatile evaluation framework for speech, audio, and music generation, it is not without its limitations. Below, we outline some areas where the toolkit could be further improved:

\noindent \textbf{Dependence on External Resources}: Many of \textit{VERSA}'s metrics require external resources, such as pre-trained models, reference datasets, or additional Python packages. The quality and diversity of these resources can impact the accuracy and fairness of evaluations, particularly for underrepresented languages and cultures. While \textit{VERSA} provides flexible configurations to accommodate various scenarios, the availability of such resources remains a bottleneck in some cases.

\noindent \textbf{Bias in Metric Design}: Despite efforts to include diverse evaluation metrics, some metrics may still reflect biases inherent in the training data or methodologies used to develop them. For example, evaluation frameworks optimized for widely spoken languages or Western music may not fully capture the nuances of less-studied languages, dialects, or musical traditions. This bias can lead to less accurate evaluations for certain domains or cultural contexts.

\noindent \textbf{Subjectivity in Perceptual Metrics}: While \textit{VERSA} incorporates automatic metrics designed to align with human subjective assessments, these metrics may not always perfectly reflect human preferences or perceptions. Human evaluations remain the gold standard for certain aspects of sound quality, naturalness, and emotional expressiveness, which automated metrics cannot fully replicate.

\noindent \textbf{Evolving Standards in Generative Models}: The rapid advancement of generative audio technologies means that new evaluation needs and metrics may arise that \textit{VERSA} does not yet support. As a result, maintaining the toolkit's relevance and adaptability will require ongoing updates and contributions from the community.

\subsection{Future Adaptation}

By accommodating a wide range of configurations, external resources, and application scenarios, we expect \textit{VERSA} to bridge the gap between human subjective assessments and automatic evaluation metrics, ensuring robust and scalable benchmarking. The adoption and usage of \textit{VERSA} are not restricted by geographic boundaries. As an open-source toolkit, it is accessible globally to researchers and practitioners, fostering international collaboration in advancing sound generation technologies. \textit{VERSA} is designed with adaptability in mind, allowing users to integrate their local resources and datasets to perform culturally and regionally relevant evaluations. 

Furthermore, \textit{VERSA} facilitates cross-border innovation by providing a standardized framework for evaluating generative audio models. This standardization reduces duplication of effort and promotes reproducibility, ensuring that advancements in sound generation and evaluation can transcend political and cultural borders. Our commitment to accessibility and inclusivity reinforces our belief in the universal potential of AI to benefit humanity without being limited by geographic, linguistic, or cultural barriers.

Looking ahead, we envision \textit{VERSA} as a key enabler for advancing the field of sound generation. Its modular design ensures adaptability to future developments, while its open-source availability fosters collaboration and community-driven enhancements. By setting a new standard for sound evaluation, \textit{VERSA} could pave the way for more transparent and effective comparisons of generative models, ultimately accelerating progress in AI-driven audio and music technologies.

\section*{Acknowledgements}
This work is supported by the Defence Science and Technology Agency (DSTA) in Singapore. We would like to thank Daniel Leong and Megan Choo for their valuable comments. Experiments of this work used the Bridges2 at PSC and Delta/DeltaAI NCSA computing systems through allocation CIS210014 from the Advanced Cyberinfrastructure Coordination Ecosystem: Services \& Support (ACCESS) program, supported by National Science Foundation grants 2138259, 2138286, 2138307, 2137603, and 2138296.


\bibliography{custom}

\appendix

\section{Example Configuration}
\label{sec:config}

\subsection{Base Configuration}

We demonstrate an example configuration file in Listing~\ref{list-config}, illustrating the ease of setup and customization within \textit{VERSA}. The configuration options are designed to cater to various user needs, offering flexibility for advanced users to tailor evaluations to specific scenarios. 

Simultaneously, default settings are provided to ensure an intuitive experience for new users, allowing them to quickly begin leveraging the toolkit without requiring extensive configuration knowledge. These thoughtful defaults strike a balance between simplicity and functionality, empowering both novice and experienced users to perform robust evaluations effortlessly.

\begin{lstlisting}[language=bash, label=list-config, ,numbers=none,caption=An example of the configuration file.]
# Example YAML config

# mcd f0 related metrics
#  -- mcd: mel cepstral distortion
#  -- f0_corr: f0 correlation
#  -- f0_rmse: f0 root mean square error
- name: mcd_f0
  f0min: 40
  f0max: 800
  mcep_shift: 5
  mcep_fftl: 1024
  mcep_dim: 39
  mcep_alpha: 0.466
  seq_mismatch_tolerance: 0.1
  power_threshold: -20
  dtw: false

# signal related metrics
# -- sir: signal to interference ratio
# -- sar: signal to artifact ratio
# -- sdr: signal to distortion ratio
# -- ci-sdr: scale-invariant signal to distortion ratio
# -- si-snri: scale-invariant signal to noise ratio improvement
- name: signal_metric

# pesq related metrics
# -- pesq: perceptual evaluation of speech quality
- name: pesq

# stoi related metrics
# -- stoi: short-time objective intelligibility
- name: stoi

# discrete speech metrics
# -- speech_bert: speech bert score
# -- speech_bleu: speech bleu score
# -- speech_token_distance: speech token distance score
- name: discrete_speech

# pseudo subjective metrics
# -- utmos: UT-MOS score
# -- dnsmos: DNS-MOS score
# -- plcmos: PLC-MOS score
- name: pseudo_mos
  predictor_types: ["utmos", "dnsmos", "plcmos", "singmos"]
  predictor_args:
    utmos:
      fs: 16000
    dnsmos:
      fs: 16000
    plcmos:
      fs: 16000

# speaker related metrics
# -- spk_similarity: speaker cosine similarity
#                    model tag can be any ESPnet-SPK huggingface repo at 
#                    https://huggingface.co/espnet
- name: speaker
  model_tag: default

# torchaudio-squim
# -- torch_squim_pesq: reference-less pesq
# -- torch_squim_stoi: reference-less stoi
# -- torch_squim_si_sdr: reference-less si-sdr
# -- torch_squim_mos: MOS score with reference
- name: squim_ref
- name: squim_no_ref

# An overall model on MOS-bench from Sheet toolkit
# More info in https://github.com/unilight/sheet/tree/main
# --sheet_ssqa: the mos prediction from sheet_ssqa
- name: sheet_ssqa

# Word error rate with OpenAI-Whisper model
# More model_tag can be from https://github.com/openai/whisper?tab=readme-ov-file#available-models-and-languages .
# The default model is `large-v3`.
# NOTE(jiatong): further aggregation are necessary for corpus-level WER/CER
# --whisper_hyp_text: the hypothesis from ESPnet ASR decoding
# --ref_text: reference text (after cleaner)
# --whisper_wer_delete: delete errors
# --whisper_wer_insert: insertion errors
# --whisper_wer_replace: replacement errors
# --whisper_wer_equal: correct matching words/character counts
# --whisper_cer_delete: delete errors
# --whisper_cer_insert: insertion errors
# --whisper_cer_replace: replacement errors
# --whisper_cer_equal: correct matching words/character counts
- name: whisper_wer
  model_tag: default
  beam_size: 5
  text_cleaner: whisper_basic

# Speech Enhancement-based Metrics
# model tag can be any ESPnet-SE huggingface repo
# -- se_sdr: the SDR from a reference speech enhancement model
# -- se_sar: the SAR from a reference speech enhancement model
# -- se_ci_sdr: the CI-SDR from a reference speech enhancement model
# -- se_si_snr: the SI-SNR from a rerference speech enhancement model
- name: se_snr
  model_tag: default
\end{lstlisting}

\subsection{Configuration Change for Model Variations}
\label{ssec:config-change}

As mentioned in Sec.\ref{ssec:supported metrics}, the configuration can be easily adjusted to accommodate different model variations. For instance, by default, we use the Rawnet-based speaker embedding~\cite{jung24c_interspeech} for speaker similarity calculation, as demonstrated in Listing~\ref{list-config-speaker}.

\begin{lstlisting}[language=bash, label=list-config-speaker,caption=An example of the speaker similarity configuration file.]
# Example Speaker Similarity YAML config

# speaker related metrics
# -- spk_similarity: speaker cosine similarity
#                    model tag can be any ESPnet-SPK huggingface repo at 
#                    https://huggingface.co/espnet
- name: speaker
  model_tag: default
\end{lstlisting}

To switch the backend speaker embedding model, you can simply modify the \texttt{model\_tag} to any supported speaker embedding model available in the ESPnet Huggingface collection.\footnote{The complete list can be found at \url{https://huggingface.co/models?other=speaker-recognition\&sort=trending\&search=espnet}.} An example is provided in Listing~\ref{list-config-speaker2}.

\begin{lstlisting}[language=bash, label=list-config-speaker2,caption=An example of the speaker similarity configuration file with a switched base model.]
# Example Speaker Similarity YAML config

# speaker related metrics
# -- spk_similarity: speaker cosine similarity
- name: speaker
  model_tag: espnet/voxcelebs12_ecapa_wavlm_joint
\end{lstlisting}

\section{Additional System Design}

\label{sec: add-system-design}

\noindent \textbf{Job-scheduling Systems}: In \textit{VERSA}'s main library, Python multi-processing is not natively supported to avoid complicating the codebase design. Instead, \textit{VERSA} achieves multi-processing through job-scheduling systems. Inspired by the approaches used in Kaldi and ESPnet~\cite{povey2011kaldi, watanabe18_interspeech}, this design enables efficient task scheduling on both computing clusters and local operating systems. To facilitate this, \textit{VERSA} includes an example script, \texttt{launch.sh}, which utilizes SLURM as the job-scheduling system to manage multi-processing tasks effectively.

\noindent \textbf{Test Protocol with Continuous Integration}: To ensure toolkit consistency and guarantee numerical stability across diverse scenarios, \textit{VERSA} is actively integrating to continuous integration testing. Unit tests are implemented for each metric, supporting a range of Python versions and external package dependencies.

\begin{table*}[t]
    \centering
    \caption{Detailed information of codec models and inference setups. }
    \resizebox{\linewidth}{!}{
    \begin{tabular}{l|c|c|c|c|c|l}
    \toprule
     Codec & kBPS & Num. Levels & Token Rate & Frame Rate & Sampling Rate & Link \\
     \midrule
       Encodec \cite{defossezhigh} & 6 & 8 & 600 & 75 & 24kHz & \tiny(\url{https://huggingface.co/facebook/encodec_24khz}) \\
       DAC \cite{kumar2024high} & 6  & 8 & 600 & 75 & 24kHz & \tiny(\url{https://github.com/descriptinc/descript-audio-codec/releases/download/0.0.4/weights_24khz.pth}) \\
       SpeechTokenizer \cite{zhang2024speechtokenizer} & 4 & 8 & 400 & 50 & 16kHz & \tiny(\url{https://huggingface.co/fnlp/AnyGPT-speech-modules/tree/main/speechtokenizer}) \\
       SQ-Codec \cite{yang2024simplespeech} & 8 & 1 & 50 & 50 & 16kHz & \tiny(\url{https://huggingface.co/Dongchao/UniAudio/resolve/main/16k_50dim_9.zip})  \\
       SoundStream (ESPnet) \cite{shi2024espnet} & 4 & 8 & 400 & 50 & 16kHz & \tiny(\url{https://huggingface.co/espnet/owsmdata_soundstream_16k_200epoch})  \\
       Speech-DAC (ESPnet) \cite{shi2024espnet} & 4 & 8 & 400 & 50 & 16kHz & \tiny(\url{https://huggingface.co/ftshijt/espnet_codec_dac_large_v1.4_360epoch}) \\
       Mimi \cite{defossez2024moshi} & 4 & 32 & 100 & 12.5 & 16kHz & \tiny(\url{https://huggingface.co/kyutai/mimi})  \\
       BigCodec \cite{xin2024bigcodec} & 1 & 1 & 80 & 80 & 16kHz & \tiny(\url{https://huggingface.co/Alethia/BigCodec/resolve/main/bigcodec.pt}) \\
       WavTokenizer~\cite{ji2024wavtokenizer} & 1 & 1 & 75 & 75 & 16kHz & \tiny(\url{https://huggingface.co/novateur/WavTokenizer-large-speech-75token})  \\
       TS3-Codec \cite{wu2024ts3} & 1 & 1 & 50 & 50 & 16kHz & -  \\
       \bottomrule
    \end{tabular}}
    \label{tab:codec_config}
\end{table*}

\begin{table*}[t]
    \centering
    \caption{\textit{VERSA} demonstration on speech coding. Codecs with streaming capacities are marked with $+$. The performance is evaluated on the LibriSpeech-test-clean set.}
    \resizebox{0.9\linewidth}{!}{
    \begin{tabular}{l|c|c|c|c|c}

    \toprule
    Codec & kbps & PESQ($\uparrow$) & UTMOS($\uparrow$) & DNSMOS (P.835)($\uparrow$) & SPK-SIM($\uparrow$)  \\
    \midrule
       Encodec\textsuperscript{+} \cite{defossezhigh}  & 6.00 & 2.77 & 3.09 & 2.96 & 0.72  \\
       DAC \cite{kumar2024high} & 6.00 & 3.40 & 3.60 & 3.16 & 0.73  \\
       SpeechTokenizer \cite{zhang2024speechtokenizer} & 4.00 & 2.62 & 3.84 & 3.17 & 0.86  \\
       SQ-Codec \cite{yang2024simplespeech} & 8.00 & \textbf{4.24} & 4.05 & 3.21 & \textbf{0.96} \\
       SoundStream (ESPnet) \cite{shi2024espnet} & 4.00 & 2.86 & 3.61 & 3.13 & 0.89  \\
       Speech-DAC (ESPnet) \cite{shi2024espnet} & 4.00 & 3.10 & 3.87 & 3.23 & 0.87  \\
       Mimi\textsuperscript{+} \cite{defossez2024moshi}  & 4.40 & 3.38 & 3.92 & 3.18 & 0.85  \\
       BigCodec \cite{xin2024bigcodec} & 1.04 & 2.68 & \textbf{4.11} & \textbf{3.26} & 0.62  \\
       WavTokenizer~\cite{ji2024wavtokenizer} & 0.98 & 1.88 & 3.77 & 3.18 & 0.60  \\
       TS3-Codec\textsuperscript{+} \cite{wu2024ts3}  & 0.85 & 2.23 & 3.84 & 3.21 & 0.70  \\
       \midrule
       Ground Truth & - & - & 4.09 & 3.18 & -  \\
       \bottomrule
    \end{tabular}}

    \label{tab:codec}
\end{table*}

\noindent \textbf{Efficient Resource Management}: Controlling external resources is a key consideration in the design of \textit{VERSA}, particularly for scenarios involving the use of external resources such as pre-trained models, datasets, and reference files. To optimize performance and reduce redundant downloads, \textit{VERSA} implements a robust cache control mechanism. This mechanism ensures that once a resource is downloaded, it is stored locally and can be reused across different metrics and evaluation runs. By centralizing resource management, \textit{VERSA} minimizes network overhead and computational costs while maintaining consistency across evaluations. 

\noindent \textbf{Functional Interface}: \textit{VERSA} is designed with versatility and ease of use in mind, offering both command-line and Python function interfaces to accommodate diverse user preferences and workflows. While the command-line interface provides a straightforward way to execute evaluations, the Pythonic API enables seamless integration into custom scripts and larger pipelines, making it particularly useful for researchers and developers looking to automate or extend evaluations. To demonstrate this flexibility, we provide an interactive Google Colab notebook showcasing how \textit{VERSA} can be effortlessly utilized within Python.\footnote{\url{https://colab.research.google.com/drive/11c0vZxbSa8invMSfqM999tI3MnyAVsOp}} This demonstration highlights the simplicity of configuring and executing evaluations programmatically, empowering users to leverage \textit{VERSA} for both standard and highly customized use cases with minimal setup.

\noindent \textbf{Community-driven Contribution}: \textit{VERSA} is designed not just as a toolkit, but as an evolving, community-driven platform that thrives on collaboration and shared innovation. To encourage and facilitate contributions, we provide clear and comprehensive guidelines for community members, available at \url{https://github.com/wavlab-speech/versa/blob/main/contributing.md}. These guidelines outline the process for proposing new features, adding metrics, reporting issues, and improving the existing codebase. By fostering an open and inclusive environment, \textit{VERSA} invites researchers, developers, and practitioners from diverse backgrounds to contribute their expertise, ensuring that the toolkit remains up-to-date and relevant for a wide range of applications. This community-driven approach not only accelerates the development of \textit{VERSA} but also ensures that it reflects the collective needs and insights of the global sound generation research community. Through active collaboration, \textit{VERSA} aims to establish itself as a dynamic and enduring resource for the advancement of sound evaluation technologies.

\begin{table*}[]
    \centering
    \caption{Detailed information of TTS models. AR and NAR stand for autoregressive and non-autoregressive architectures. }
        \resizebox{\linewidth}{!}{\begin{tabular}{l|c|c|l}
    \toprule
    Model & Feature Type & Model Type & Pre-trained Model Link  \\
    \midrule
    ESPnet-TTS \cite{hayashi2021espnet2} & Continuous & NAR & \tiny(\url{https://huggingface.co/espnet/kan-bayashi_ljspeech_vits})  \\
    ESPnet-SpeechLM & Discrete & AR & \tiny(\url{https://huggingface.co/espnet/speechlm\_tts\_v1
}) \\
    ChatTTS \cite{chattss2024} & Discrete & AR & \tiny(\url{https://huggingface.co/2Noise/ChatTTS}) \\
    CosyVoice  \cite{du2024cosyvoice} & Discrete & AR & \tiny(\url{https://huggingface.co/model-scope/CosyVoice-300M}) \\
    EmotiVoice \cite{emotivoice2024} & Continuous & NAR & \tiny(\url{https://www.modelscope.cn/syq163/outputs.git}) \\
    MeloTTS  \cite{zhao2024melo} & Continuous & NAR & \tiny(\url{https://huggingface.co/myshell-ai/MeloTTS-English}) \\
    Parler-TTS \cite{lacombe-etal-2024-parler-tts}  & Discrete & AR & \tiny(\url{https://huggingface.co/parler-tts/parler-tts-mini-v1}) \\
    WhisperSpeech \cite{whisperspeech2024} & Discrete & AR & \tiny(\url{https://huggingface.co/WhisperSpeech/WhisperSpeech/blob/main/t2s-v1.95-small-8lang.model}) \\
    VallE-X \cite{zhang2023speak} & Discrete & AR \& NAR & \tiny(\url{https://huggingface.co/Plachta/VALL-E-X/resolve/main/vallex-checkpoint.pt}) \\
    Vall-E 2 \cite{chen2024vall} & Discrete & AR \& NAR & \tiny(\url{https://huggingface.co/amphion/valle}) \\
    NaturalSpeech2 \cite{shennaturalspeech} & Continuous & NAR & \tiny(\url{https://huggingface.co/amphion/naturalspeech2_libritts}) \\
    \bottomrule
    \end{tabular}}

    \label{tab:tts_model}
\end{table*}

\begin{table*}[]
    \centering
    \caption{\textit{VERSA} demonstration on TTS. We use $\dagger$ for TTS systems with a fixed speaker.}
    \resizebox{\linewidth}{!}{
    \begin{tabular}{l|c|c|c|c|c|c|c|c|c}
    \toprule
    Model & UTMOS($\uparrow$) & DNSMOS (P.835)($\uparrow$) & SHEET-SSQA($\uparrow$) & PLCMOS($\uparrow$) & SPK-SIM($\uparrow$) & TS-PESQ($\uparrow$) & TS-SI-SNR($\uparrow$) & TS-MOS($\uparrow$) & Whisper-WER($\downarrow$) \\
    \midrule
    ESPnet-TTS \cite{hayashi2021espnet2} & 4.06 & 3.05 & 4.02 & 4.51 & - & 3.61 & 26.55 & 4.25 & 5.01 \\
    ESPnet-SpeechLM & 4.05 & 3.24 & 4.24 & 4.44 & \textbf{0.55} & 3.58 & 23.94 & 4.33 & 3.12 \\
    ChatTTS$\dagger$  \cite{chattss2024} & 3.52 & 3.24 & 4.17 & 4.56 & - & 3.42  & 23.63 & 4.28 &  7.09 \\
    CosyVoice$\dagger$  \cite{du2024cosyvoice} & 4.15 & \textbf{3.25} & 4.40 & 4.51 & 0.51 & 3.59 & 22.43 & \textbf{4.41} & 4.95 \\
    EmotiVoice$\dagger$  \cite{emotivoice2024} & \textbf{4.22} & 3.04 & \textbf{4.49} & 4.68 & - & \textbf{4.14} & \textbf{28.61} & 4.15
 & \textbf{3.10} \\
    MeloTTS$\dagger$  \cite{zhao2024melo} & 3.80 & 3.15 & 3.99 & \textbf{4.71} & - & 3.65 & 27.77 & 4.28
 & 4.64 \\
    Parler-TTS$\dagger$  \cite{lacombe-etal-2024-parler-tts} & 3.83 & 3.16 & 4.34 & 4.55 & - & 3.87 & 25.78 & 4.27 & 4.66 \\
    WhisperSpeech$\dagger$  \cite{whisperspeech2024} & 4.06 & \textbf{3.25} & 4.36 & 4.66 & - & 3.92 & 27.43 & 4.33 & 13.45 \\
    VallE-X \cite{zhang2023speak} & 3.38 & \textbf{3.25} & 3.64 & 4.45 & 0.35 & 3.58 & 21.42 & 4.34 &  27.33 \\
    Vall-E 2 \cite{chen2024vall} & 3.65 & 2.88 & 3.84 & 3.67 & 0.46 & 3.31 & 18.74 & 4.12 & 27.83 \\
    NaturalSpeech2 \cite{shennaturalspeech} & 2.35 & 2.24 & 3.16 & 4.27 & 0.32 & 2.27 & 10.85 & 3.62 & 4.62 \\
       \midrule
       Ground Truth & 4.09 & 3.18 & 4.35 & 4.16 & - & 3.59 & 24.01 & 4.53 & 2.59 \\
       \bottomrule
    \end{tabular}}

    \label{tab:tts}
\end{table*}

\section{Experimental Demonstration}
\label{sec:exp}

This appendix includes the experimental results for the demonstration discussed in Section~\ref{sec:demo}. 

\subsection{Codec Evaluation Demonstration}
\label{ssec:codec}

We demonstrate the usage of \textit{VERSA} for speech coding evaluation with nine open-source models, including Encodec~\cite{defossezhigh}, DAC~\cite{kumar2024high}, Mimi~\cite{defossez2024moshi}, BigCodec~\cite{xin2024bigcodec}, SpeechTokenizer~\cite{zhang2024speechtokenizer}, Wavtokenizer~\cite{ji2024wavtokenizer}, SQ-Codec~\cite{yang2024simplespeech}, TS3-Codec~\cite{wu2024ts3}, and several ESPnet-Codec public checkpoints~\cite{shi2024espnet}. We use the test-clean set in Librispeech for evaluation~\cite{panayotov2015librispeech}.

For all selected models, we utilize their public checkpoints along with the corresponding released inference scripts. Additionally, for multi-level codec models, we limit the codec levels to ensure alignment within a comparable bitrate range. Detailed inference hyperparameter configurations and the corresponding results are provided in Table~\ref{tab:codec_config}.

Table~\ref{tab:codec} presents the evaluation results using the \textit{VERSA} minimum installation. While detailed model discussions are beyond the scope of this paper, \textit{VERSA} offers a versatile perspective for analyzing models and assessing their advantages. This capability significantly reduces the effort both model developers and end-users need to select codecs tailored to specific downstream applications. For example, the high pseudo-MOS scores (UTMOS and DNSMOS) observed in BigCodec indicate superior synthesis quality. However, the lower signal-level metrics (e.g., SDR and PESQ) suggest that the low bitrate leads to a substantial loss of signal detail.

While the current investigation serves as a demonstration with a small portion of the supported metrics in \textit{VERSA}, we aim to conduct a more comprehensive analysis in future work, ensuring broader coverage across various aspects of sound generation.

\begin{table*}[]
    \centering
    \caption{\textit{VERSA} demonstration on speech enhancement. The performance is evaluated on the Voicebank-DEMAND test set.}
    \resizebox{0.9\linewidth}{!}{
    \begin{tabular}{l|c|c|c|c|c}
    \toprule
    Model & UTMOS($\uparrow$) & DNSMOS (P.835)($\uparrow$) & SHEET-SSQA($\uparrow$) & PESQ($\uparrow$) & STOI($\uparrow$) \\
    \midrule
    Noisy & 3.11 & 2.52 & 3.57 & 1.97 & 0.92  \\
    \midrule
       MetricGAN+~\cite{fu21_interspeech} & 3.63 & 2.95 & 3.91 & \textbf{3.15} & 0.93 \\
       MTL-Mimic~\cite{bagchi2018spectral} & \textbf{3.86} & \textbf{3.03} & \textbf{4.06} & 3.01 & \textbf{0.95} \\
       SepFormer~\cite{subakan2021attention} & 3.68 & 2.87 & 3.63 & 3.13 & 0.93 \\
       \bottomrule
    \end{tabular}}

    \label{tab:se}
\end{table*}
\begin{table*}[]
    \centering
    \caption{Detailed pre-trained model links for the singing voice synthesis demonstration.  }
        \resizebox{\linewidth}{!}{\begin{tabular}{l|l}
    \toprule
    Model &  Pre-trained Model Link  \\
    \midrule
    RNN~\cite{shi2021sequence} & \url{https://huggingface.co/espnet/opencpop_naive_rnn_dp} \\
    XiaoiceSing~\cite{lu20c_interspeech} & \url{https://huggingface.co/espnet/opencpop_xiaoice} \\
    DiffSinger~\cite{liu2022diffsinger} & \url{https://github.com/MoonInTheRiver/DiffSinger/releases/download/pretrain-model/0228_opencpop_ds100_rel.zip} \\
    VISinger~\cite{zhang2022visinger} & \url{https://huggingface.co/espnet/opencpop_visinger} \\
    VISInger2~\cite{zhang23e_interspeech} & \url{https://huggingface.co/espnet/opencpop_visinger2} \\
    TokSing~\cite{wu24q_interspeech} &  \url{https://huggingface.co/espnet/opencpop_svs2_toksing_pretrain} \\
    VISinger2+~\cite{yu2024visinger2} &  \url{https://huggingface.co/yifengyu/svs_train_visinger2plus_mert_raw_phn_None_zh_200epoch} \\
    \bottomrule
    \end{tabular}}

    \label{tab:svs_model}
\end{table*}

\begin{table*}[]
    \centering
    \caption{\textit{VERSA} demonstration on singing voice synthesis. The performance is evaluated on the Opencpop test set.}
    \resizebox{0.9\linewidth}{!}{
    \begin{tabular}{l|c|c|c|c|c|c}
    \toprule
    Model & MCD($\downarrow$) & F0-RMSE($\downarrow$) & F0-CORR($\uparrow$) & SHEET-SSQA($\uparrow$) & SingMOS($\uparrow$) & SPK-SIM($\uparrow$) \\
    \midrule
    RNN~\cite{shi2021sequence} & 9.52 & 86.40 & 0.51 & 3.38 & 3.64 & 0.58\\
    XiaoiceSing~\cite{lu20c_interspeech} & 9.54 & 69.05 & 0.57 & 3.13 & 3.53 & 0.57 \\
    DiffSinger~\cite{liu2022diffsinger} & 8.22 & 63.84 & 0.61 & 4.15 & \textbf{4.25} & 0.71 \\
    VISinger~\cite{zhang2022visinger} & 7.91 & 60.20 & 0.62 & 4.12 & 4.13 & \textbf{0.76} \\
    VISinger2~\cite{zhang23e_interspeech} & \textbf{7.89} & 59.53 & 0.62 & 4.19 & 4.21 & 0.62 \\
    TokSing~\cite{wu24q_interspeech} & 8.10 & 61.59 & 0.62 & 3.73 & 3.86 & 0.67 \\
    VISinger2+~\cite{yu2024visinger2} & 8.73 & \textbf{55.26} & \textbf{0.66} & \textbf{4.32} & 4.22 & 0.68\\
       \midrule
       Ground Truth & - & - & - & 4.49 & 4.79 & - \\
       \bottomrule
    \end{tabular}}

    \label{tab:svs}
\end{table*}
\begin{table*}[]
    \centering
    \caption{Detailed pre-trained model links for the music generation demonstration.}
        \resizebox{0.9\linewidth}{!}{
    \begin{tabular}{l|l}
    \toprule
    Model & Pre-trained Model Link \\
    \midrule
       AudioLDM2 \cite{liu2023audioldm} & \url{https://huggingface.co/cvssp/audioldm2-music} \\
       MusicGen \cite{copet2024simple} & \url{https://huggingface.co/facebook/musicgen-large} \\
       MusicLDM \cite{chen2024musicldm} & \url{https://github.com/RetroCirce/MusicLDM?tab=readme-ov-file\#step-4-run-musicldm} \\
       Riffusion-v1\cite{Forsgren_Martiros_2022} & \url{https://huggingface.co/riffusion/riffusion-model-v1} \\
       Stable-Audio-Open \cite{stable-audio} & \url{https://huggingface.co/stabilityai/stable-audio-open-1.0} \\
       \bottomrule
    \end{tabular}}

    \label{tab:music_model}
\end{table*}

\begin{table*}[]
    \centering
    \caption{\textit{VERSA} demonstration on music generation. }
    \resizebox{\linewidth}{!}{
    \begin{tabular}{l|c|c|c|c}
    \toprule
    Model & FAD-LAION-CLAP($\downarrow$) & KID-LAION-CLAP($\downarrow$) & Density-LAION-CLAP($\uparrow$) & Coverage-LAION-CLAP($\uparrow$) \\
    \midrule
       AudioLDM2 \cite{liu2023audioldm} & 47.97 & 2.05 &  0.04 &  0.0058  \\
       MusicGen-large \cite{copet2024simple} & \textbf{20.06} & \textbf{0.49} & 0.07 & 0.0514 \\
       MusicLDM \cite{chen2024musicldm} & 40.46 & 1.62 & 0.07 & 0.0118 \\
       Riffusion-v1\cite{Forsgren_Martiros_2022} & 52.58 & 2.63 & 0.01 & 0.0024 \\
       StableAudioOpen \cite{stable-audio} & 20.99 & 0.62 & \textbf{0.08} & \textbf{0.0674} \\
       \bottomrule
    \end{tabular}}

    \label{tab:music}
\end{table*}

\subsection{TTS Evaluation Demonstration}
\label{ssec:tts}

Similar to the codec demonstration, we utilize a range of open-source TTS models for the TTS evaluation, including ESPnet-TTS~\cite{hayashi2021espnet2}\footnote{We use the VITS model~\cite{kim2021conditional} trained on LJSpeech~\cite{ljspeech17}.}, ESPnet-SpeechLM,\footnote{\url{https://github.com/espnet/espnet/tree/speechlm}} ChatTTS~\cite{chattss2024}, CosyVoice~\cite{du2024cosyvoice}, EmotiVoice~\cite{emotivoice2024}, MeloTTS~\cite{zhao2024melo}, Parler-TTS~\cite{lacombe-etal-2024-parler-tts}, WhisperSpeech~\cite{whisperspeech2024}, VallE-X~\cite{zhang2023speak}, and Vall-E 2~\cite{chen2024vall}. Specifically, we use the common LibriSpeech test-clean set for evaluation. For zero-shot TTS systems, we provide speaker prompts by randomly selecting a prompt segment from the target speaker. Table~\ref{tab:tts_model} provides a high-level categorization of feature types—discrete or continuous acoustic features—and model types, namely autoregressive or non-autoregressive. Inference is performed using the open-source pipelines provided by the model developers, with no additional hyperparameter tuning.

Some models, such as Parler-TTS and CosyVoice, struggled to interpret capitalized input text from LibriSpeech. To address this, we optionally applied a case-recovery preprocessing step for those models. Importantly, no post-filtering was applied to autoregressive models utilizing discrete tokens.

In Table~\ref{tab:tts}, we present the performance of various TTS models. Since these models have been optimized using different frameworks and datasets, deriving definitive scientific conclusions is challenging. However, two potential implications can be drawn in this context:
\begin{itemize}
    \item There remains room for improvement in zero-shot TTS performance compared to fixed-speaker TTS systems.
    \item Discrete token-based autoregressive modeling continues to be a popular approach. However, without sufficient post-filtering, it struggles to achieve stability comparable to that of non-autoregressive methods.
\end{itemize}

Building on this demonstration, we anticipate a more comprehensive analysis utilizing all 54 speech-related metrics in \textit{VERSA} for existing TTS models in our future work. This analysis would further explore the current state of TTS development and address the challenges associated with existing evaluation metrics.


\subsection{Speech Enhancement Evaluation Demonstration}
\label{ssec:se}
We demonstrate the speech enhancement evaluation with three public available models, including SepFormer~\cite{subakan2021attention}, MetricGAN~\cite{fu21_interspeech}, and MTL-MIMIC~\cite{bagchi2018spectral}. We use VoiceBank-DEMAND as the test set~\cite{valentini2017noisy}.

The evaluation of speech enhancement is demonstrated in Table~\ref{tab:se}. While this paper focuses on six specific metrics, we anticipate broader adoption of the task by leveraging the diverse set of 65 metrics available in \textit{VERSA}.

\subsection{Singing Synthesis Evaluation Demonstration}
\label{ssec:svs}

For the singing voice synthesis demonstration, we select a range of open-source singing voice synthesis models, including Naive-RNN~\cite{shi2021sequence}, XiaoiceSing~\cite{lu20c_interspeech}, VISinger~\cite{zhang2022visinger}, VISinger2~\cite{zhang23e_interspeech}, TokSing~\cite{wu24q_interspeech}, VISinger2+~\cite{yu2024visinger2}, and DiffSinger~\cite{liu2022diffsinger}.\footnote{The DiffSinger is trained with its own open-source repository~\cite{liu2022diffsinger}, where other models are trained with Muskit-ESPnet~\cite{shi22d_interspeech, wu2024muskits}.} The links for pre-trained models are listed in Table~\ref{tab:svs_model}.

We demonstrate the usage of \textit{VERSA} in singing voice synthesis in Table~\ref{tab:svs}. Compared to speech quality assessment, singing voice-related analysis usually utilizes subjective evaluation due to the limited resources available. With the support of \textit{VERSA}, we additionally support SHEET-SSQA~\cite{huang2024mos} and SingMOS~\cite{tang2024singmos}, which consider singing voice synthesis, aligning a range of open-source singing synthesis models.

\subsection{Music Generation Evaluation Demonstration}
\label{ssec:music}

We utilize five models for music generation evaluation, including AudioLDM2~\cite{audioldm2-2024taslp}, MusicGen~\cite{copet2024simple}, MusicLDM~\cite{chen2024musicldm}, Riffusion-v1~\cite{Forsgren_Martiros_2022}, StableAudioOpen~\cite{stable-audio}. Table~\ref{tab:music_model} shows the pre-trained model links used for the demonstration. To generate music from each model, we utilized text prompts from MusicCaps, a dataset comprising expert-annotated captions of 30-second YouTube clips~\cite{agostinelli2023musiclm}. We employed the rewritten, quality-neutral prompts provided in \cite{gui2024adapting}. Additionally, ChatGPT was used to filter out prompts referring to non-instrumental music or those not typically associated with the text-to-music generation, such as tutorials or environmental sounds.

For evaluation, we utilize the final layer embedding of LAION-CLAP (Music pre-trained version) to compute the metrics~\cite{laionclap2023}.\footnote{\url{https://github.com/LAION-AI/CLAP}} FMA-small dataset is used as the reference dataset for distributional analysis~\cite{fma_dataset}.

Table~\ref{tab:music} presents the evaluation results obtained using \textit{VERSA} for music generation tasks. The results highlight the strengths and unique characteristics of different models. Specifically, MusicGen-large demonstrates robust capabilities in generating coherent and high-quality songs, showcasing its strength in musical structure and tonal consistency. On the other hand, StableAudioOpen excels in diversity, producing a wide range of musical styles and compositions. These results underscore the utility of \textit{VERSA} in providing a detailed, multi-faceted evaluation of music generative models, enabling researchers to identify and compare model-specific strengths.

\end{document}